\documentclass[
twocolumn, 
amssymb, 
prl]{revtex4-1} % single column

\usepackage{amsmath}
\usepackage{hyperref}% add hypertext capabilities
\hypersetup{
    colorlinks=true,
    linkcolor=blue,
    filecolor=magenta,      
    urlcolor=cyan,
}
\usepackage{graphicx}
\usepackage{epstopdf}
\usepackage{caption}
\usepackage{subcaption}
\usepackage{color}
\usepackage{dcolumn}% Align table columns on decimal point
\usepackage{bm}% bold math\
\usepackage{ulem}

\begin{document}

% \title{Bayesian modeling and optimization for physical processes}
\title{Bayesian optimization of a free-electron laser}
% \title{We need to decide on a title....}
% \title{Squanching with Bayes}

\author{J. Duris}
\author{D. Kennedy}
\author{A. Hanuka}
\author{J. Shtalenkova}
\author{A. Edelen}
\affiliation{SLAC National Accelerator Laboratory, Menlo Park, CA 94025, USA}
\author{M. McIntire}
\author{S. Ermon}
\affiliation{Stanford University Department of Computer Science, Stanford, CA 94305, USA}
\author{A. Egger}
\author{T. Cope}
\author{D. Ratner}
\affiliation{SLAC National Accelerator Laboratory, Menlo Park, CA 94025, USA}

\date{\today}

\begin{abstract}

% before cutting to proper length for prl (797 / 600 characters)
The Linac Coherent Light Source changes configurations multiple times per day, necessitating fast tuning strategies to reduce setup time for successive experiments. To this end, we employ a Bayesian approach to transport optics tuning to optimize groups of quadrupole magnets. We use a Gaussian process to provide a probabilistic model of the machine response with respect to control parameters from a modest number of samples. Subsequent samples are selected during optimization using a statistical test combining the model prediction and uncertainty. The model parameters are fit from archived scans, and correlations between devices are added from a simple beam transport model. The result is a sample-efficient optimization routine, which we show significantly outperforms existing optimizers.

% ... upon checking a few published PRLs, it seems we're fine

% % after cutting to proper length for prl (599 / 600 characters)
% The Linac Coherent Light Source changes configurations multiple times per day, necessitating fast tuning strategies to reduce setup time for successive experiments. To this end, we apply Bayesian optimization to transport optics tuning to optimize groups of quadrupole magnets, using a Gaussian process to provide a probabilistic model of the machine response. The model’s parameters are fit from archived scans and correlations between devices are added from a beam transport model. The result is a sample-efficient optimization routine, which we show significantly outperforms existing optimizers.

\end{abstract}

% \pacs{}
% 41.60.Cr - Free-electron lasers
% 42.55.Vc - X- and γ-ray lasers

\maketitle

%================================================================
% \section{Introduction and Motivation}

% problem description (tuning, time spent...)
% history of other methods
% NOTE: This opening should be rewritten to be made more general depending on target audience.
% Note: this section presently being rewritten + need to reduce accelerator-specific jargon and repetition

% \cite{LCLS2002,emma:lcls}
Modern large-scale scientific experiments can have complicated operational requirements with performance degraded by errors in controls and dependencies on drifting or random variables. A prime example of this is the Linac Coherent Light Source (LCLS) \cite{emma:lcls}, an x-ray free electron laser (FEL) user facility that supports a wide array of scientific experiments. At LCLS, skilled human operators tune dozens of control parameters on-the-fly to achieve custom photon beam characteristics, and this process cuts into valuable time allocated for each user experiment. Model-independent optimizers can help automate tuning, with successful demonstrations using simplex \cite{simplex,tomin:ocelot}, extremum seeking \cite{ES,es_rf_tuning_paper}, and robust conjugate direction search \cite{rcds,rcds_demo}. However, these methods require a large number of expensive acquisitions and can become stuck in local optima. To improve the efficiency of optimization beyond these methods, a model of the system is necessary \cite{good_regulator}; in this work, we use Bayesian optimization with Gaussian process (GP) models during live tuning of LCLS. First, we use archived data to estimate the length scales of tuning parameters in the model. Second, we show that adding physics-inspired correlations between parameters further speeds convergence. Finally, we discuss possible directions for improvement.
% [Move to end of GP section]First, we estimate the GP kernel parameters from data, using both an intuitive approach of estimating basis functions and Bayesian inference to estimate covariance parameters. Second, we show that physics-inspired kernel parameters further speed convergence. Finally, we discuss natural extensions for future work.
%================================================================
% \section{Bayesian optimization and Gaussian Processes}
% Intro to Bayesian opt and GPs

% Bayes theorem which allows the incorporation of prior knowledge or evidence $E$ to guide the development of a probabilistic model $M$ of the system given evidence: $p(M|E) \propto p(E|M) p(M)$. 
Bayesian optimization is a sample efficient and gradient-free approach to global optimization of black-box functions with noisy outputs \cite{mockus1975, freitas:bayes_opt_review, freitas:tutorial}. 
This efficiency comes from application of Bayes theorem to incorporate prior knowledge and previous steps to maximize the value of each new measurement. 
Numerical optimization of an acquisition function, incorporating the model's expectation and uncertainty, guides the selection of each new point to sample, giving Bayesian optimization the advantage of balancing exploration with exploitation. 
Prior knowledge can improve the efficiency of the optimization and constrain the search in low signal-to-noise states. 
Bayesian optimization has been applied to a fast growing number of domains; for example, resource prospecting \cite{kriging:origins,kriging:water}, active policy search for reinforcement learning \cite{bayes-policy-search}, hyper parameter tuning \cite{bayes-hyperparam-tuning}, experimental control \cite{yager2019} etc.

Bayesian optimization requires a probabilistic model providing estimates and uncertainties of a target or objective function. A Gaussian process (GP) is a popular choice as it is a non-parametric regressor which calculates probability densities over a space of functions \cite{rw:gpml_book}. Whereas a Gaussian distribution is characterized by a mean and covariance $y \sim {\sc N} (\boldsymbol{\mu},\Sigma)$, a Gaussian \textit{process} is determined by mean and covariance \textit{functions}: $f(\boldsymbol{x}) \sim {\sc GP} (m(\boldsymbol{x}),k(\boldsymbol{x},\boldsymbol{x'}))$, where $\boldsymbol{x}$ are possible inputs to the objective.
The mean $m(\boldsymbol{x})$ is a predetermined function encoding prior understanding of the objective. This mean could be a fit to data but is typically set to zero. 
The covariance, or kernel, function $k(\boldsymbol{x},\boldsymbol{x'})$ describes the similarity between pairs of points $\boldsymbol{x}$ and $\boldsymbol{x'}$. As a non-parametric model, a GP is constructed directly from the training instances themselves, allowing the model complexity to grow with observations and adapt to previously unexplored regions of the feature space.
% Hyperparameters are chosen by maximizing the marginal likelihood for a set of observed data.
% Maximizing the GP marginal likelihood of a set of observed data guides the kernel functional form and hyperparameter selection. 
% Maximizing the GP marginal likelihood of a set of observed data guides kernel and hyperparameter selection, and modeling noise with the kernel avoids overfitting.
There are many Gaussian process codes available \cite{GPy,GPFlow,GPML,PyMC3}, and many of these packages employ various techniques to limit computation times for large data sets. In our case, we use an online GP model \cite{mitch:spwogp} interfaced to LCLS via the Ocelot optimization framework \cite{tomin:ocelot}. The online model saves frequent computations to speed subsequent predictions.

%================================================================
% \section{Kernel parameters from archived data}

% Benefit: exploration/exploitation trade-off ==> efficient exploration
% Method: basic isotropic GP (just length scales)
% Length scales from archive

In this Letter we apply Gaussian process optimization to the problem of maximizing the LCLS x-ray FEL pulse energy. The FEL instability is a collective effect so performance depends strongly on the current density and therefore beam size. A strong focusing quadrupole magnet system maintains a small beam size to maximize the pulse energy produced. The FEL pulse energy is therefore a function of quadrupole magnet strengths (field integral in kiloGauss, kG) and the input electron beam parameters. Since these beam parameters drift throughout the day and are generally hard to measure, LCLS operators perform tuning scans to optimize the quadrupole strengths many times per day.

We model the FEL dependence on quadrupoles with a Gaussian process surrogate. Prior information guides the selection of the kernel and its hyperparameters. We approach this process in two ways: first, from an intuitive view of basis functions, and second in a principled way using Bayesian model selection. 
An attractive feature of Gaussian process modeling is the interpretability of the kernel's functional form. 
The prediction can be viewed as a convolution of measured data with a kernel function response to measured data points (analogous to a Green's function approach);
%, and the effect is to decompose the system response into basis functions with maximal inner product with the system being modeled. 
thus, kernel functions should be derived from basis functions that look like the system being modeled. 
We exploit this insight here, explaining the interpretation of each kernel hyperparameter, and then show that the basis function approach also maximizes the GP marginal likelihood.

Motivated by the observation that the FEL pulse energy response to quadrupole magnets looks bell-shaped, we choose the popular radial basis function (RBF) kernel, $k_{\rm RBF}(\boldsymbol{x}, \boldsymbol{x}') = \sigma_f^2 \exp(-\frac{1}{2}(\boldsymbol{x} - \boldsymbol{x}')^{\rm T} \Sigma (\boldsymbol{x} -  \boldsymbol{x}'))$. Here, $\sigma_f^2$ is the covariance function amplitude and $\Sigma$ is a diagonal matrix of inverse square length scales. This kernel encodes the expectation that for smooth functions, nearby pairs of points are more similar than distant points. The length scales set the distance over which the function changes in each dimension. The amplitude parameter captures the variance of the target function values with respect to variations in the inputs and therefore determines the prior prediction uncertainty far from any sampled points. Functions which are less smooth may be better modeled by the Matern kernel, whereas periodic functions are best treated with a periodic kernel. Kernels may be added or multiplied together to yield new, more expressive kernels \cite{duvenaud:kernel_composition,sun:nkn} and neural network warping functions may be applied to the inputs before passing to a kernel \cite{manifoldgp,wilson:dkl}. Modeling noisy targets can be achieved by adding a Gaussian noise kernel $k_{\rm noise}(\boldsymbol{x}, \boldsymbol{x}') = \sigma_{n}^2 \delta(\boldsymbol{x} - \boldsymbol{x}')$, where $\sigma_{n}^2$ is the noise variance parameter, and $\delta$ is the Dirac delta function. The noise parameter models the variance of the prediction at a sampled point.  

\begin{figure} [htbp]
\centering
\begin{subfigure}[]{0.23\textwidth}
\centering
    \includegraphics[width= 0.93\textwidth] {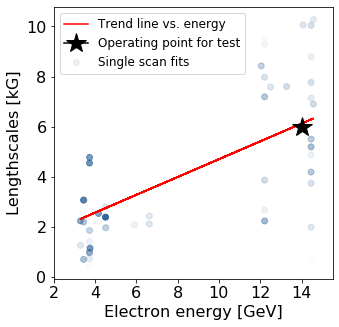}
    \caption{Fit Lengthscales}
    \label{fig:1b}
   \end{subfigure}
\begin{subfigure}[]{0.23\textwidth}
\centering
 \includegraphics[width= 0.93\textwidth]{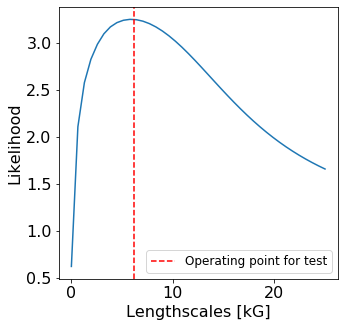}
   \caption{Marginal likelihood}
   \label{fig:1a}
\end{subfigure}
\\
\begin{subfigure}[]{0.23\textwidth}
\centering
    \includegraphics[width= 0.93\textwidth] {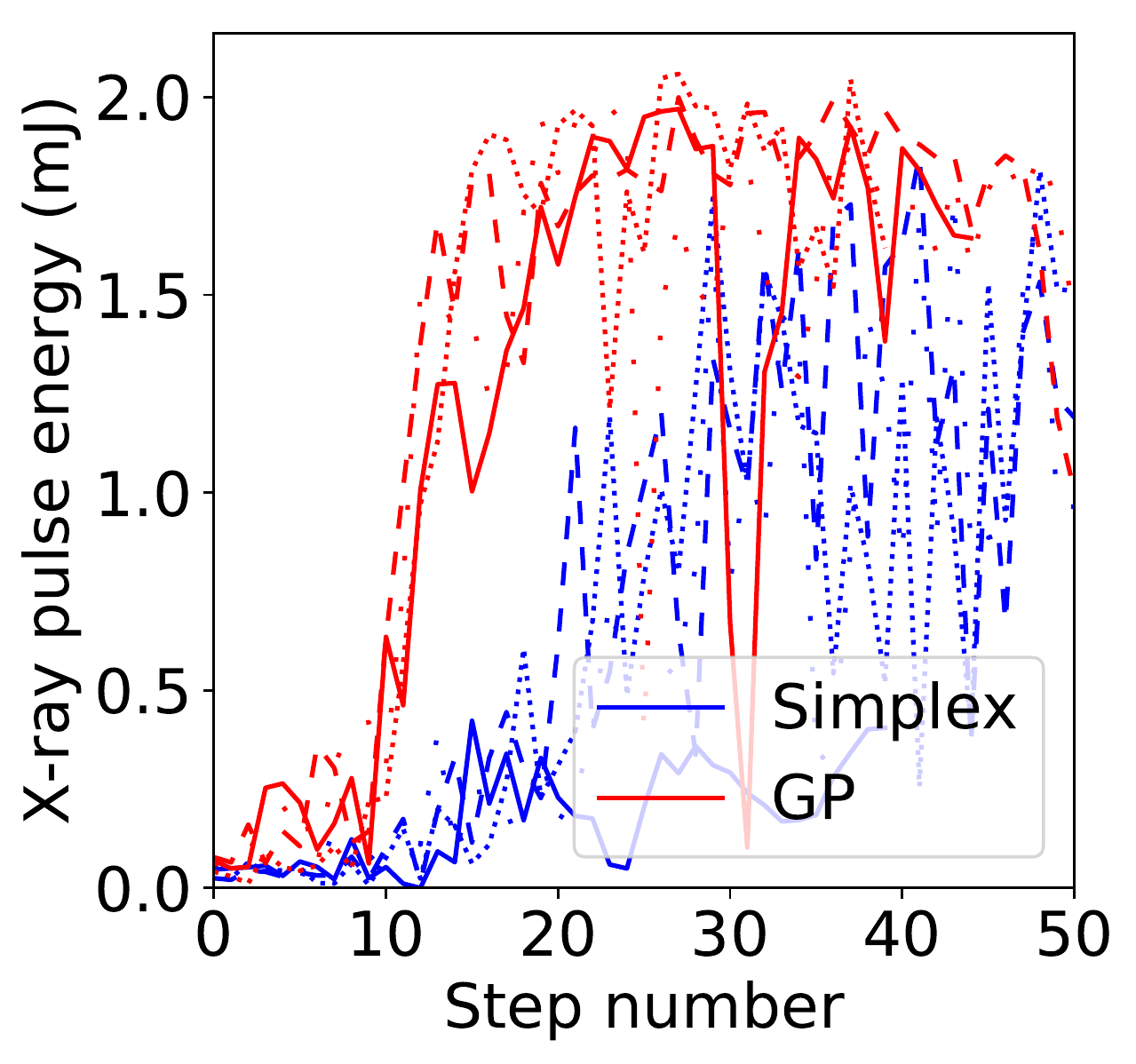}
    \caption{Live 12 quadrupole optimization}
    \label{fig:1c}
\end{subfigure}
\begin{subfigure}[]{0.23\textwidth}
\centering
    \includegraphics[width= 0.93\textwidth] {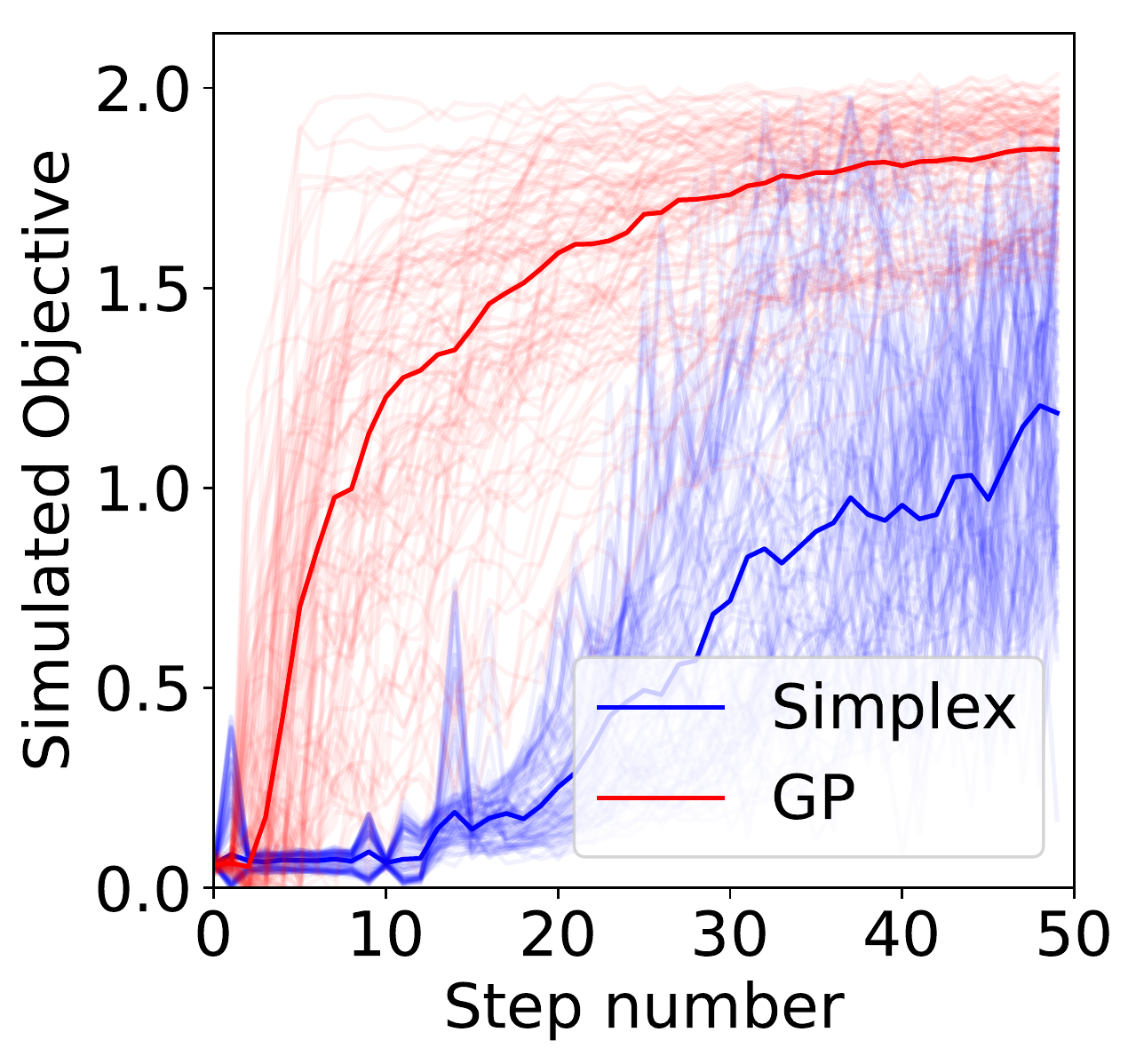}
    \caption{Simulated 12 quadrupole optimization}
    \label{fig:1d}
\end{subfigure}
   \caption{(\subref{fig:1b}) Fit length scales in kG vs beam energy in GeV for one quadrupole. The red line is a linear fit to the length scale with points weighted by uncertainty of the individual point estimates (more transparent points have less certainty). The black star represents the Gaussian fit, which was used in the optimization. (\subref{fig:1a}) Example of a GP marginal likelihood fit vs length scale at 14 GeV. (\subref{fig:1c}) Comparison of optimization of FEL pulse energy over 12 quadrupole magnets for Bayesian optimization vs Nelder-Mead simplex optimizer. Each scan was performed 4 times with identical starting conditions, shown with different dashing. Each step corresponds to approximately 3 seconds of beam time.
   (\subref{fig:1d}) Simulations using the beam matrix model for the conditions of (c). 100 individual scans for each method, with means shown by thick lines, are consistent with measurements.}
\label{fig:archivefits}
\end{figure}

The RBF length scales depend on the electron beam energy because the focusing strength of a quadrupole magnet depends on the ratio of the electron energy to the magnetic field strength (i.e. the rigidity). 
Moreover, the beamsize response varies between each quadrupole even at a single energy.
We therefore determine RBF length scales for each quadrupole independently as a function of electron beam energy.
We first approach the estimation of these length scales with 1D Gaussian fits to historical tuning scans.
The black star in Fig.~\ref{fig:1b} shows the average of  many such fit lengths to scans with electron beam energies near 14~GeV for a particular quadrupole.
We also estimate the RBF length scales from maximization of the marginal likelihoods of GP fits to the archive data. 
An example of this procedure for a scan at 14~GeV is shown in Fig.~\ref{fig:1a}, where we can see that 50\% deviations from the optimal length scale decreases the marginal likelihood by merely 10\%--providing some tolerance to errors in the parameter determination. Figure~\ref{fig:1b} shows the maximum likelihood length scales for various scans as a function of electron beam energy. 
The red line is a linear fit to the length scales with points weighted by widths of the marginal likelihoods (effectively the credibilities) for each individual point estimate. The resulting trend is then used to construct RBF kernels for optimization, and we observe reasonable agreement with our length scale estimate from the Gaussian fits. We repeat this procedure to determine length scale trends as a function of electron energy for each quadrupole.

Online optimization proceeds by first measuring the initial state of the machine and then initializing the Gaussian process model with the appropriate kernel and first measured point. The GP provides a probabilistic surrogate model for the machine, and an upper confidence bound (UCB) acquisition function \cite{ucb} is constructed from the GP prediction mean and variance. The point maximizing the UCB function is chosen as the next measurement which is then acquired and added to the GP, finishing one step through the optimization process. The optimization continues in this way until reaching a time limit or a target performance.

Figure~\ref{fig:1c} shows results from live optimization of the FEL pulse energy simultaneously on 12 quadrupoles. In this example, Bayesian optimization (red curves) is approximately 4 times faster than the standard Nelder-Mead simplex algorithm (blue curves), and reaches a higher optimum. The different lines for each algorithm correspond to scans with identical starting conditions. 

We also compare simplex and Bayesian optimization in a simulation environment. Ideally we would use physics codes such as elegant \cite{elegant} and Genesis \cite{Genesis} to model the transport and FEL behavior, however due to mismatch between the codes and measured performance, as well as computational expense for each simulation, we instead fit a beam transport model as described below. Though the beam transport model does not capture the full complexity of the real machine, it allows us to compare the relative performance of simplex and Bayesian optimization with a simulated objective function, which we find consistent with live scans (Fig.~\ref{fig:1d}).

%================================================================
% \section{Physics-informed kernel correlations}

\begin{figure} [htbp]
\centering
\begin{subfigure}[]{0.235\textwidth}
\centering
 \includegraphics[width= 0.99\textwidth]{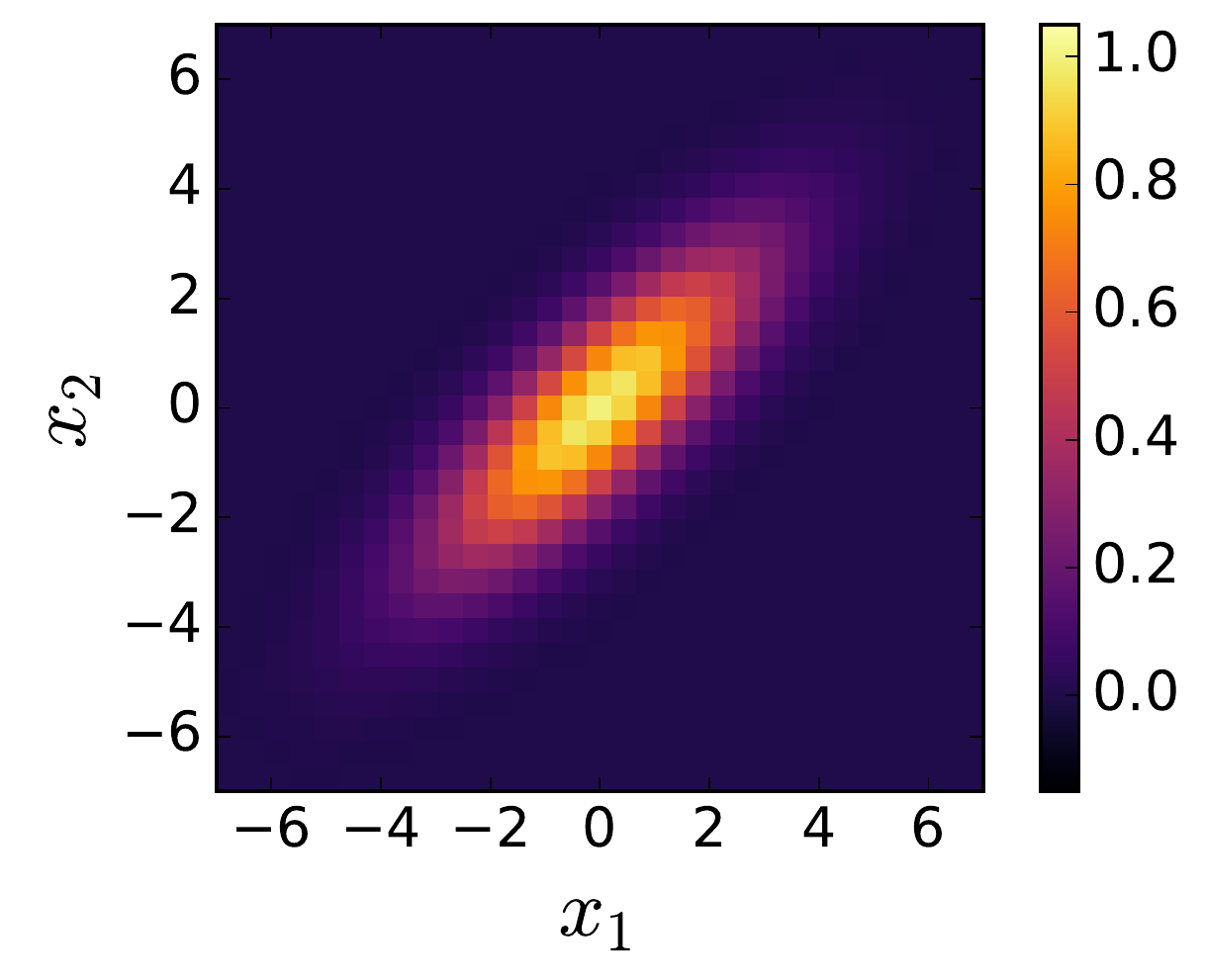}
   \caption{Ground truth}
   \label{fig:2a}
   \end{subfigure}
\begin{subfigure}[]{0.235\textwidth}
\centering
    \includegraphics[width= 0.99\textwidth] {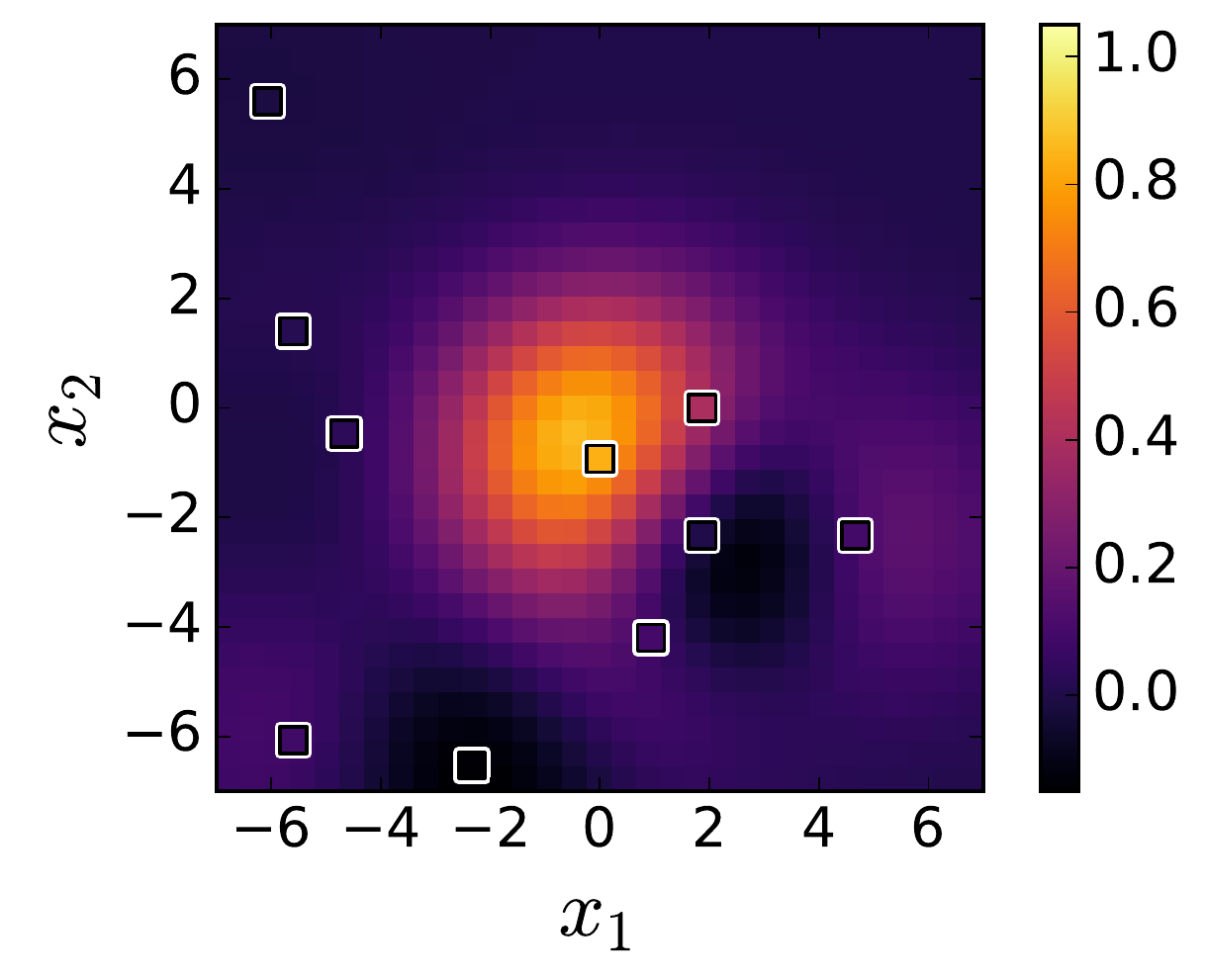}
    \caption{Isotropic kernel}
    \label{fig:2b}
\end{subfigure}
\\
\begin{subfigure}[]{0.235\textwidth}
\centering
    \includegraphics[width= 0.99\textwidth] {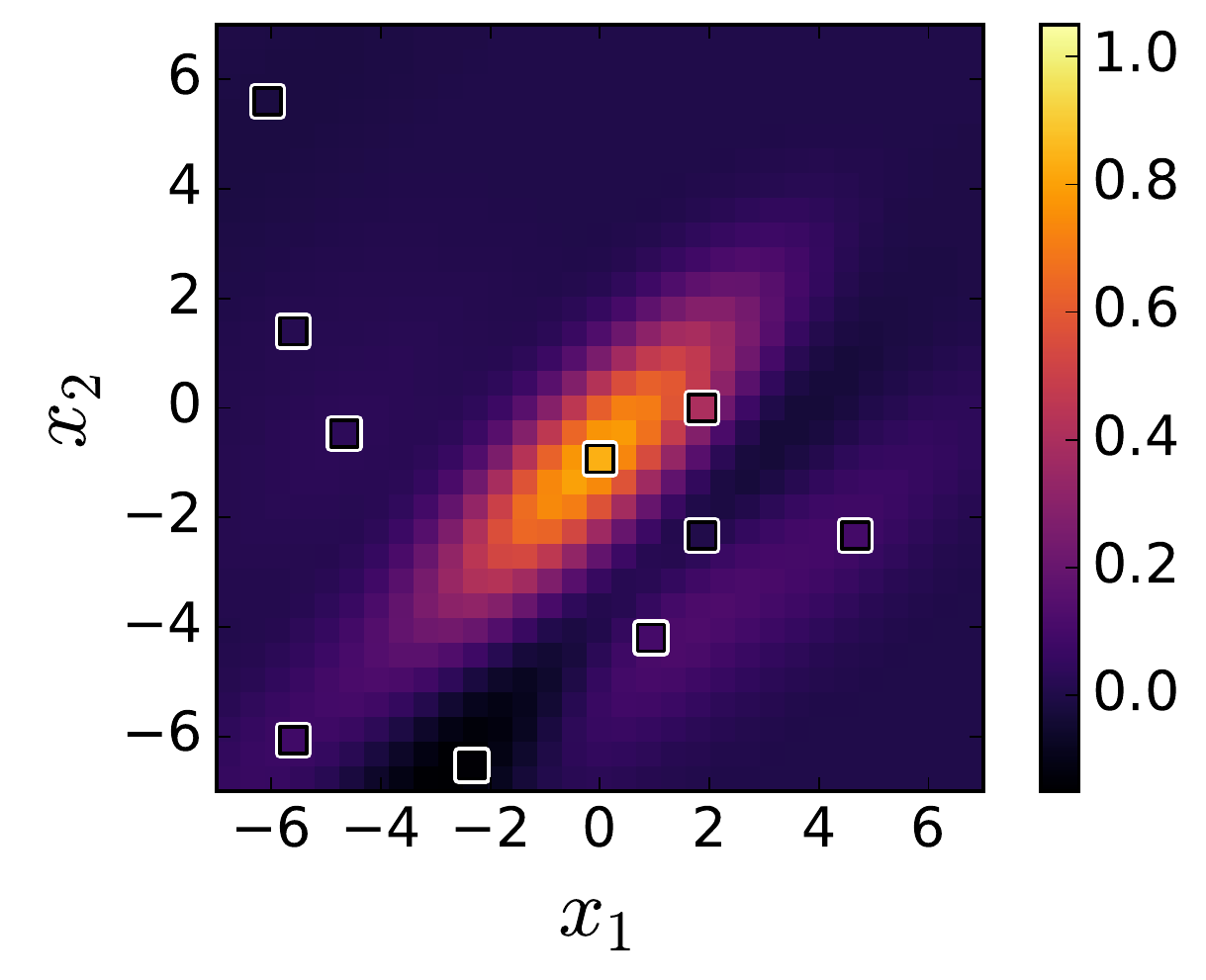}
    \caption{Correlated kernel}
    \label{fig:2c}
\end{subfigure}
\begin{subfigure}[]{0.235\textwidth}
\centering
    \includegraphics[width= 0.83\textwidth] {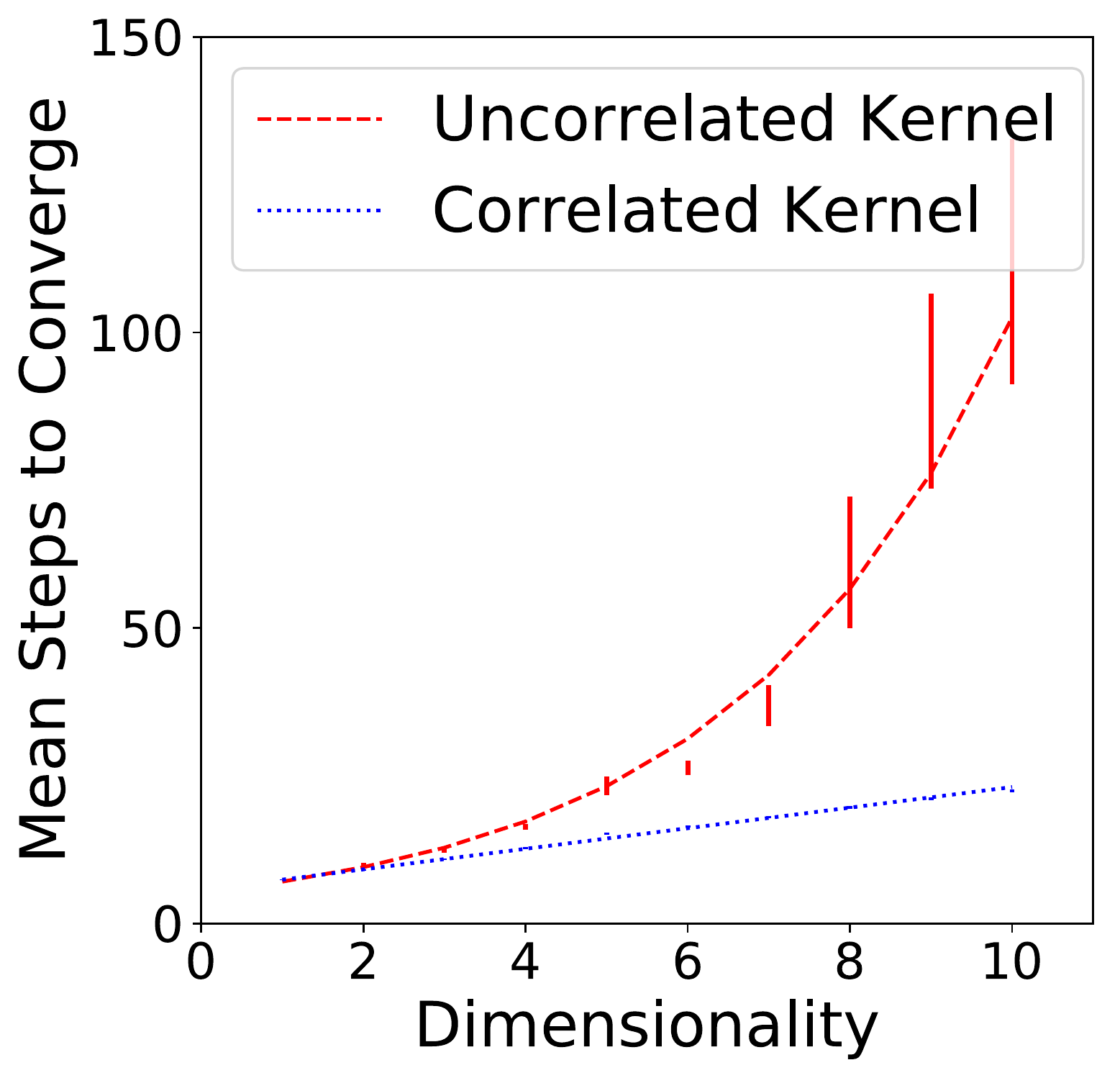}
    \caption{Convergence tests}
    \label{fig:2d}
\end{subfigure}
   \caption{(\subref{fig:2a}) Ground truth of a test function with unit slice widths and correlation coefficient of 0.8. (\subref{fig:2b}) GP regression with isotropic kernel on noisy samples from ground truth. (\subref{fig:2c}) GP regression with correlated kernel on identical samples as b. (\subref{fig:2d}) Bayesian optimization convergence tests on a correlated ground truth with and without kernel correlations. Each bar shows the standard error about the mean for 100 trials. The correlated GP kernel (blue linear fit) performs as well as optimization of an isotropic ground truth with an isotropic GP kernel, growing linearly with the number of dimensions. Steps to convergence with mismatched kernel grows exponentially (red exponential fit).}
\label{fig:correlations}
\end{figure}

% item3) might be good to compare a likelihood ratio to compare the "goodness" of the models
Correlated kernels become advantageous when a system's response is to one input depends strongly on one or more of the other inputs as in Figure~\ref{fig:2a}. Figure~\ref{fig:2b} shows a GP regression on noisy samples (RMS noise is 10\% of the signal peak) from a correlated ground truth (Fig.~\ref{fig:2a}) with an isotropic kernel, while Figure~\ref{fig:2c} shows a GP regression on the \textit{same} samples but with a kernel sharing the same correlation as the ground truth ($\rho=0.8$). The latter model is more representative of the system. To demonstrate the effectiveness of a correlated model for regulating the system, we perform Bayesian optimization with and without kernel correlations for various dimensional spaces with nearest-neighbor correlation coefficients of $\rho_{i,i+1}=0.5$. Each point in Figure~\ref{fig:2d} shows the average number of steps to achieve $>90\%$ of the ground truth peak amplitude for 100 runs starting at a random position such that the starting signal-to-noise ratio is unity.
The relative efficiency of the correlated kernel grows exponentially with the number of dimensions, making it attractive for high-dimensional optimization at accelerators.
%{\color{red} The GP optimization with correlated kernel outperforms the isotropic one because the area of the hyper-ellipse in the correlated space grows much more slowly than that of the hyper-sphere that encloses it. Thus, modeling correlations between various elements effectively reduces the search volume.}

% \begin{figure}
% \includegraphics[width=0.48\textwidth]{fig2}
% \caption{a) Ground truth with unit slice widths and correlation coefficient of 0.7 b) GP regression with isotropic kernel. c) GP regression with correlated kernel. d) Optimization convergence tests with and without correlations.}
% \label{fig:correlations}
% \end{figure}

\begin{figure} [htbp]
\centering
\begin{subfigure}[]{0.23\textwidth}
\centering
 \includegraphics[width= 0.99\textwidth]{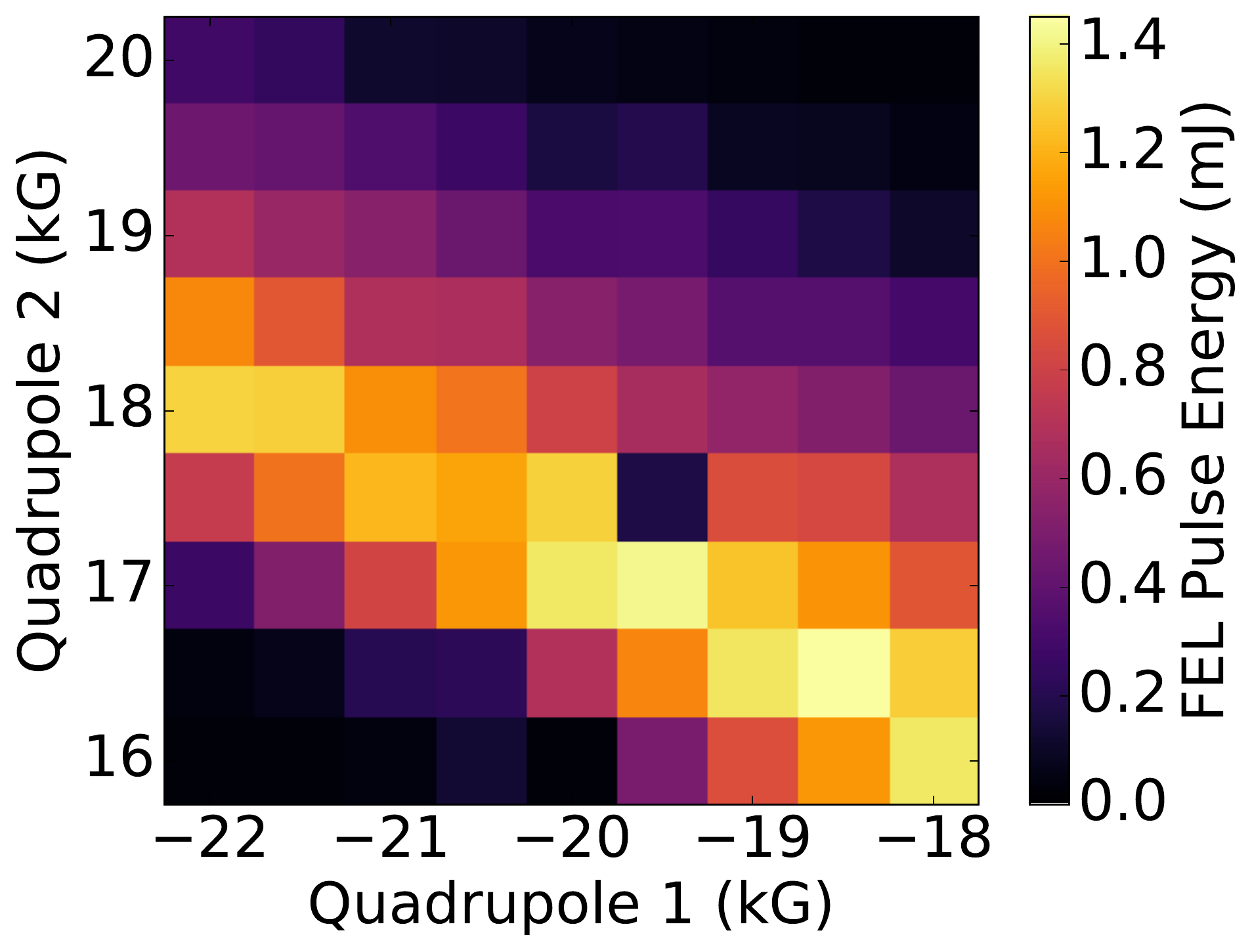}
   \caption{Measured pulse energy}
   \label{fig:3a}
   \end{subfigure}
\begin{subfigure}[]{0.23\textwidth}
\centering
    \includegraphics[width= 0.99\textwidth] {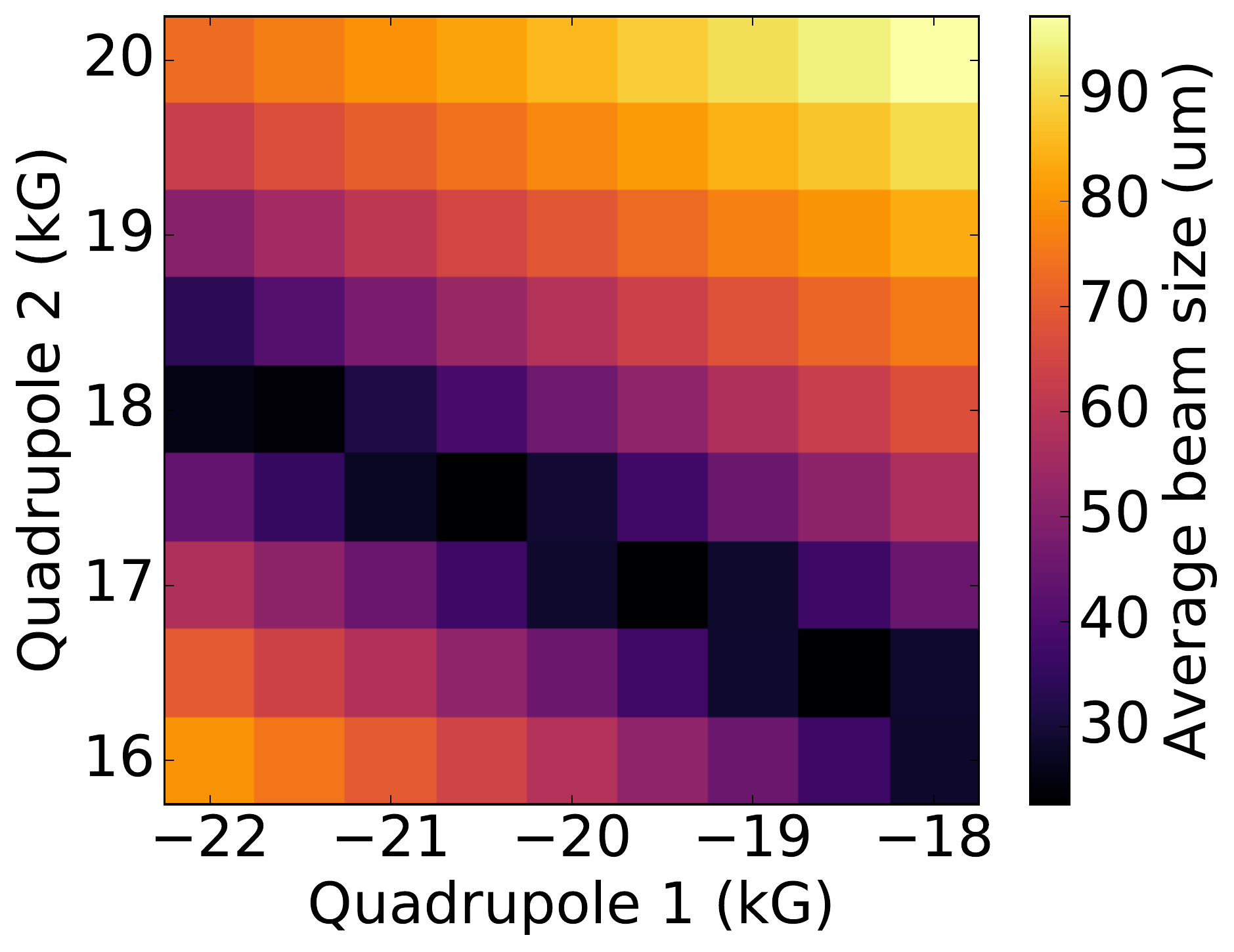}
    \caption{Modeled beam size}
    \label{fig:3b}
\end{subfigure}
\\
\begin{subfigure}[]{0.23\textwidth}
\centering
    \includegraphics[width= 0.9\textwidth] {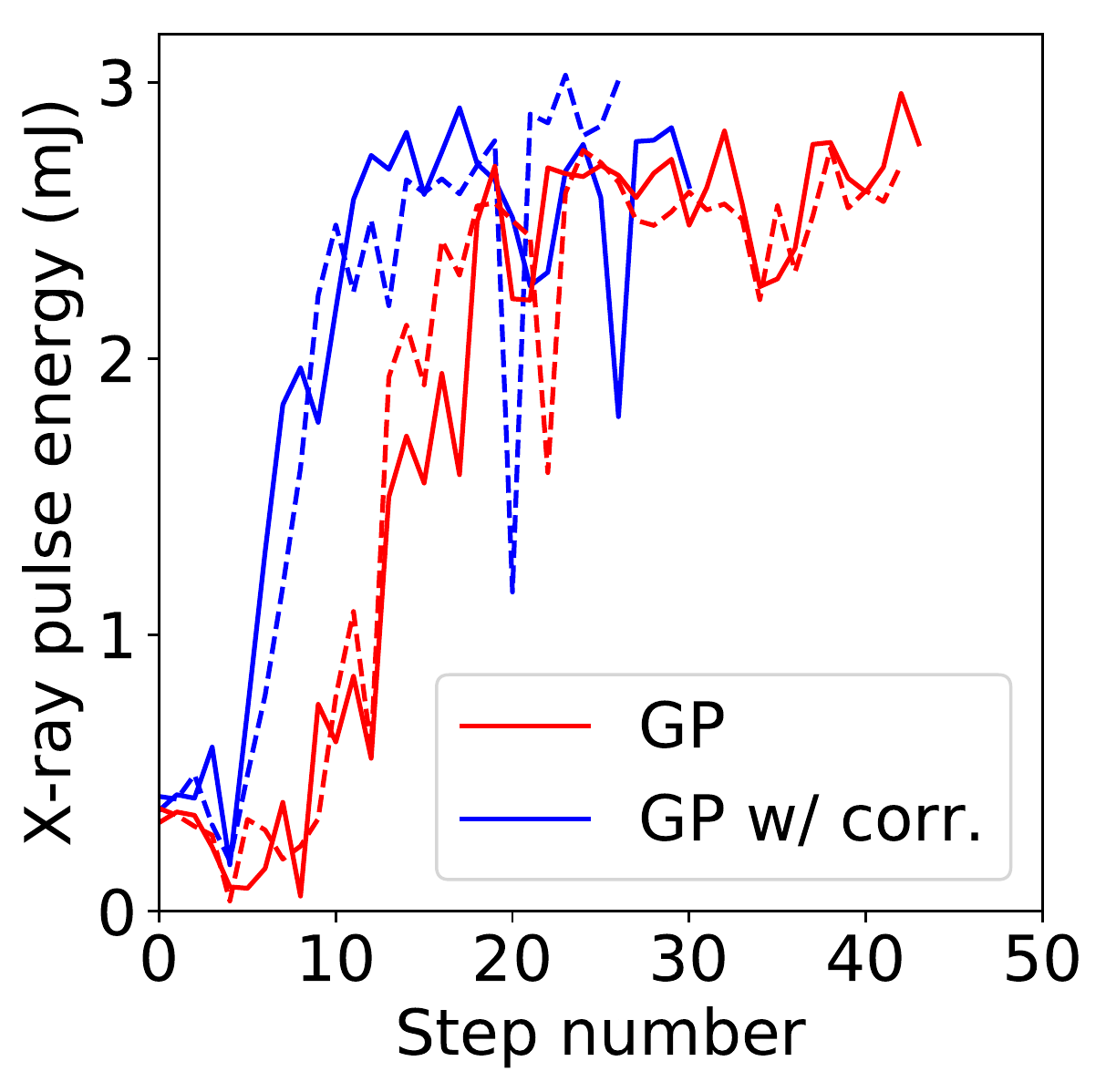}
    \caption{Live 4 quadrupole optimization}
    \label{fig:3c}
\end{subfigure}
\begin{subfigure}[]{0.23\textwidth}
\centering
    \includegraphics[width= 0.9\textwidth] {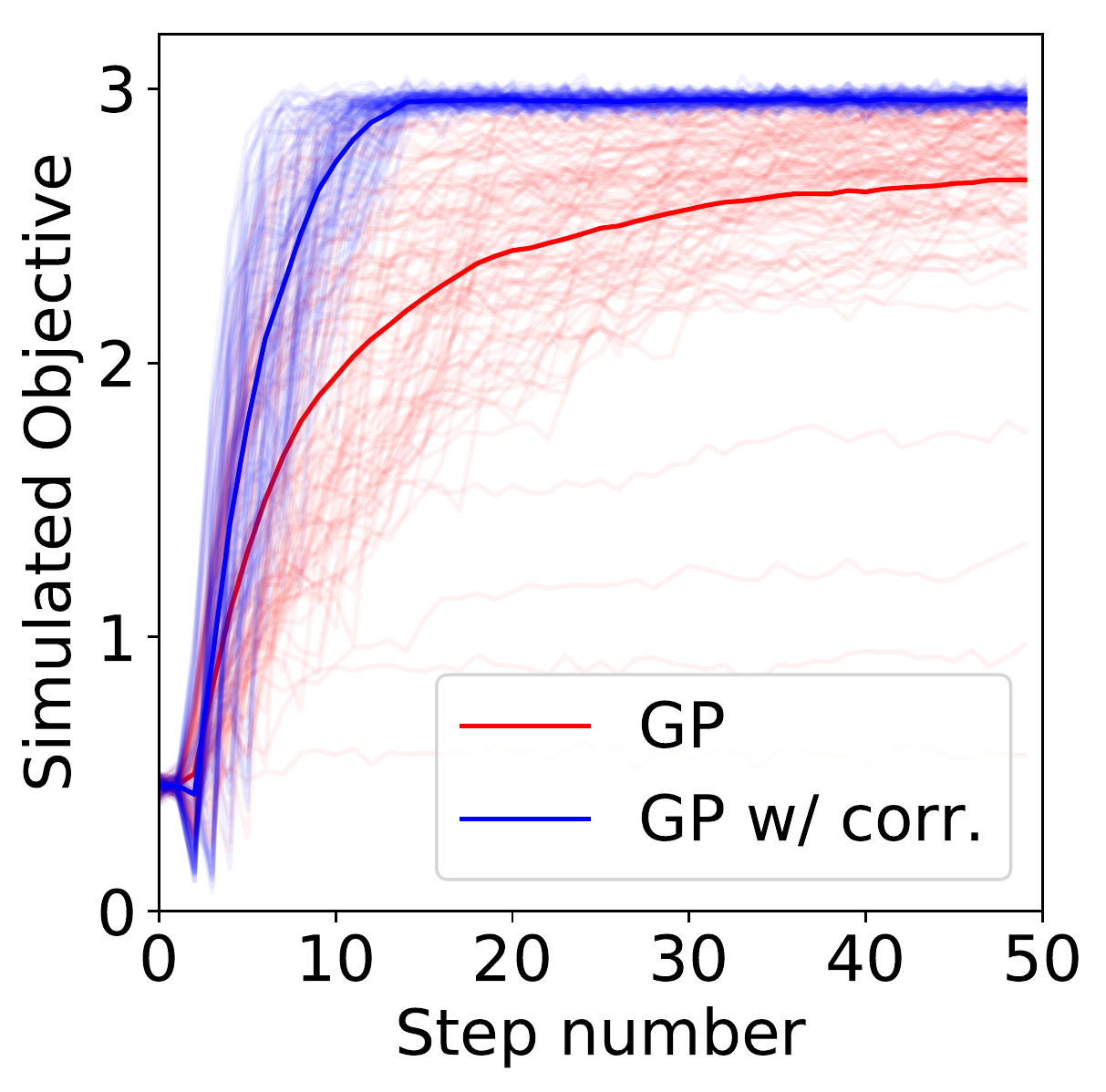}
    \caption{Simulated 4 quadrupole optimization}
    \label{fig:3d}
\end{subfigure}
   \caption{(\subref{fig:3a}) Average FEL pulse energy vs two adjacent quadrupoles. (\subref{fig:3b}) Average modeled electron beam size in the undulator vs the same quadrupoles preserves the correlations. (\subref{fig:3c}) Optimization test for 4 quadrupoles: GP vs GP with correlations. Each scan was performed twice with identical starting conditions, shown with different dashing. (\subref{fig:3d}) Simulations using a beam matrix model for the conditions of \subref{fig:3c}. 100 individual scans for each method, with means shown by thick lines, are consistent with measurements.}
\label{fig:physicskernel}
\end{figure}

FEL quadrupole optimization is an example of a highly correlated system. Strong focusing in charged particle transport relies on a series of oppositely polarized quadrupole fields \cite{wiedemann}. Each quadrupole focuses in one transverse plane while defocusing in the orthogonal plane, and repeated application of alternate focusing/defocusing results in net focusing in both planes. Increasing the strength of one quadrupole field necessitates increasing the strength of the next quadrupole (with opposite sign) to achieve net focusing, resulting in negative correlations between nearby focusing elements. Figure~\ref{fig:3a} shows the average FEL pulse energy response to variation of two adjacent quadrupoles in a matching section just upstream of the FEL undulator magnets.
Similar correlations exist between all pairs of quadrupoles.

% beam phys model = correlations
Ideally we would calculate correlations from maximum likelihood fits to the archive data, as done for the length scales.  However, due to sparsity in the sampled data, we instead calculate correlations from a beam physics transport model. The FEL pulse energy (denoted as $U$) is a correlation-preserving function of the transverse area of the beam, $\sigma^2$, averaged along the interaction, with $\log U \propto \langle \sigma^2 \rangle^{-1/3}$. \cite{zhirong:fel_book}. As a consequence, any correlated response of the quadrupole magnets with respect to beam size preserves correlations with respect to the FEL energy.

% what is the beam physics model
To model the average beam size in the undulator line as a function of quadrupole magnets, we estimate the Twiss, or focusing, parameters of the electron beam. Realistic Twiss parameters may be inferred from measurements of the beam size throughout a drift or in response to varying focusing \cite{emittance_measurements}. Alternatively, we can use the premise that maximized FEL performance implies approximately matched Twiss parameters in the undulator line. Starting from these initial conditions, we then use linear transport matrices calculated from live quadrupole values \cite{wiedemann} to calculate the beam's size along the undulator for any arbitrary deviation in quadrupole strength. The result shown in Figure \ref{fig:3b} is a modeled beam size that matches the correlation to the FEL pulse energy (Fig. \ref{fig:3a}).

% how do we get correlations from this model
To find correlations from the beam physics model, we approximate the beam size about the match by the Hessian $H_{i,j}=\partial_{x_i} \partial_{x_j} \langle \sigma \rangle$, where $i$ and $j$ iterate over all pairs of quadrupoles. Correlations are extracted by identifying the diagonal elements as the functional curvature along each slice $C=\sqrt{\mathrm{diag}(H)}$ and factoring these out of the Hessian $h = C^{-1} H C^{-1}$.  

% now we need to combine them
We now need to combine the correlations with the length scales in the RBF kernel. In principle, the Hessian can also be used to calculate the length scales. However, the post-saturation FEL process is computationally expensive to model and further complicated by leaked dispersion and orbit jitter so characteristic length scales are difficult to compute on the fly. As a result, we combine the modeled correlations with length scales measured from fits to archive data as described earlier. A diagonal matrix of inverse slice lengths $S_{i,j}=\sigma_{i}^{-1}\delta_{i,j}$ determined from archived scans are combined with the precision matrix in the kernel, $\Sigma=S h S$, to yield a kernel with slice lengths from archived scans and correlations from the simulated beam transport model. With a more sophisticated model that incorporates a wider range of effects, it should be possible to train hyperparameters entirely from simulations.

% testing
We tested optimization with GPs built with and without correlations for four adjacent matching quadrupole magnets located in the transport between the accelerator and the FEL undulators. The results, presented in Figure~\ref{fig:3c}, show that the correlated GP model outperforms the uncorrelated GP. %, which in turn  outperforms the current Nelder-Mead simplex algorithm. 
Furthermore, we find consistency of these results with simulations based on correlations from the beam transport model as shown in Figure~\ref{fig:3d}. 
It is interesting to note that the Bayesian optimizer maximized the FEL with a similar number of steps for the 12 quadrupole case (Fig.~\ref{fig:1c}), seemingly in contradiction to Figure~\ref{fig:2d}. The similarity is due to the fact that to leading order (assuming a monoenergetic beam and linear optics), only four quadrupole magnets are needed to match the four Twiss parameters into the undulator line. However, optimizing more quadrupoles further increases the FEL pulse energy by reducing chromatic effects which suppress FEL gain by increasing the electron beam emittance.
While four quadrupoles can recover a significant fraction of peak performance, in practice operators typically cycle through subsets of all of the controllable quadrupoles. 
% Future work will focus on controlling larger sets of input devices, in which we expect to see increasingly superior performance when using GP with correlations.
With the ever increasing number of input dimensions or controls, the GP with correlations is expected to perform exponentially faster than without.
%Future work will expand the GP with correlations to control larger sets of quadrupoles.
% scan all 24 tuning quadrupoles. 

% % could possibly skip over this if we didn't truncate significantly
% % this may also lead to a non-local kernel? linear kernel => predict?
% If the A degenerate or nearly degenerate scaled inverse correlation matrix $C^{-1}$ may possibly lead to a correspondingly degenerate Gramm or covariance matrix $K$ which could prevent inversion during the GP prediction. Furthermore, near degeneracy may have large scaled lengths, which suggest predictive power far from the sampled point. In order to limit this predictive power, we calculate the eigenvalues of this matrix and truncate off-diagonal elements until the maximum eigenvalue is 2 or 3. In this way, we inform the optimizer of correlations while limiting numerical instability.

%================================================================
% \section{Summary \& Future outlooks}

In this letter we showed that online Bayesian optimization with length scales estimated from archived historical data can tune the LCLS FEL pulse energy more efficiently than the current state-of-the-art Nelder-Mead simplex algorithm. Moreover, we showed that adding physics-informed correlations, trained from beam optics models, further improves tuning efficiency. The latter effect was demonstrated with four control parameters, and the improvement is expected to grow dramatically as the number of dimensions increases when controlling more variables such as RF variables or undulator strengths. The flexibility of the GP enables training from archived data, simulations, and physical models.

We expect Bayesian optimization will become a standard tuning method for LCLS-II. Future improvements will further improve efficiency and operational ability. This Letter focused on the most time-consuming task of tuning quadrupoles, but we expect additional applications to other accelerator and beamline tasks, e.g. reducing bandwidth, optimizing taper \cite{juhao:taperopt}, focusing and alignment, etc. Adding `safety' constraints to the acquisition function can guide the exploration while ensuring operational requirements are met (for instance, avoiding transient drops in FEL pulse energy or keeping losses low) \cite{kirschner:swissfel}. More expressive GPs, for example deep kernel learning \cite{DKL} or deep GPs \cite{Deep-GP} can learn more complicated functions and extract additional value from historical data. In our example, we exploited the physics of strong focusing to learn correlations between parameters. Physics abounds with well verified mathematical models and incorporating additional physics knowledge into the prior, either explicitly in the formulation of the kernel function or via additional training with simulation can have a dramatic effect on optimization efficiency. %<Joe do we want to add direct learning of underlying dynamics + papers along these lines?>. Joe's response: We could but it's not necessary. Can you give an example?

%kernel spectral learning

% Probably need to introduce FELs and LCLS. This is getting complicated...
% Unlike many applications of machine learning, physics abounds with well verified mathematical models. Physical theories are often far more computationally efficient and generalizable (interpolate and extrapolate well) compared to fits to observed data in a local domain. In the case of FEL and beam dynamics, we can leverage beam dynamics models to describe relationships observed in data.

%We plan to calculate the entire kernel including length scales and correlations from a more robust online FEL model.

%%%%%%%%%%%%%%%%%%%%%%%%%%%%%%%%%%%%%%%%%
\begin{acknowledgments}
The authors are grateful to Hugo Slepicka and Sergey Tomin for their help with the details of the Ocelot-optimizer code and to Hugo Slepicka and Ahmed Osman for adding helpful LCLS-specific functions to Ocelot-optimizer.
This work was supported by the Department of Energy, Laboratory Directed Research and Development program at SLAC National Accelerator Laboratory, under contract DE-AC02-76SF00515.

\end{acknowledgments}

%%%%%%%%%%%%%%%%%%%%%%%%%%%%%%%%%%%%%%%
% \bibliographystyle{unsrt}
% \bibliography{references2}

\end{document}